\documentclass{article}

\usepackage{amsmath}
\usepackage{amssymb}
\usepackage{graphicx}
\usepackage{color}

\begin{document}

\begin{flushright}
{KEK-TH-1551}
\end{flushright}
\vskip 1cm

\begin{center}
{\LARGE{\bf
The Einstein equation of state\\ as the Clausius relation\vspace{10pt}\\ with an entropy production}
}
\vskip 1cm

\renewcommand{\thefootnote}{\fnsymbol{footnote}}
{\large 
Kengo Shimada\footnote[1]{e-mail address: skengo@post.kek.jp},
Susumu Okazawa\footnote[2]{e-mail address: okazawas@post.kek.jp}
and Satoshi Iso\footnote[3]{e-mail address: satoshi.iso@kek.jp}
}
\vskip 0.5cm

{\it 
High Energy Accelerator Research Organization (KEK)  \\
and  \\
The Graduate University for Advanced Studies (SOKENDAI), \\
Oho 1-1, Tsukuba, Ibaraki 305-0801, Japan
}
\end{center}

\vskip 1cm
\begin{center}
\begin{bf}
Abstract
\end{bf}
\end{center}

We give a modified derivation of the Einstein equation of state
by considering the Clausius relation $T\delta S-\delta N =\delta Q$ 
on a null hypersurface with a non-vanishing expansion 
($\theta \neq 0$), i.e. not in the {\it equilibrium}. 
The derivation corresponds to choosing a specific observer to the
hypersurface, and such a generalization gives a hint how we can improve the 
original derivation by Jacobson. We also give an interpretation
of the thermodynamic relation based on the Noether charge method.

\renewcommand{\thefootnote}{\arabic{footnote}}
\section{Introduction}

\hspace{16pt}Black hole thermodynamics has been extensively investigated as a hint towards understanding the microscopic structure of space-time. 
The first law of black hole thermodynamics is usually given as a relation of thermodynamic quantities between two equilibrium black holes in the Einstein-Hilbert theory of gravity \cite{Bardeen:1973gs}.
It is generalized by Wald to any diffeomorphism invariant gravity theories with higher derivative terms {\cite{Wald:1993nt,Jacobson:1993vj,Iyer:1994ys},
and the derivation of the first law  essentially uses the fact that the black hole horizon is a Killing horizon.
The first law is also investigated in a physical process of  throwing matter into a black hole \cite{Wald,Gao:2001ut} in which process there is no  Killing vector.
The physical process version was proved only in a subclass of theories including $F(R)$ and Lanczos-Lovelock gravities \cite{Jacobson:1995uq,Chatterjee:2011wj,Kolekar:2012tq}.

The notion of entropy associated with the area of the horizon and the thermodynamics is also generalized to the Rindler horizon \cite{Unruh:1976db}, 
which strongly suggests a deep connection with the emergence of space-time and its thermodynamic origin.
In \cite{Jacobson:1995ab},
Jacobson proposed to derive the Einstein equation of motion starting from the thermodynamic relation,
i.e. the Clausius relation between a change of entropy and an energy flux 
\begin{align}
T \delta S = \delta Q
\label{Clausius}
\end{align}
across the local causal horizon (LCH). 
As depicted in Figure \ref{LCH},
he first considered a point $p$ in an $n$-dimensional space-time on an $(n-2)$-dimensional space-like hypersurface ${\cal P}$ and future-directed null vectors $K^\mu$ on ${\cal P}$ perpendicular to the surface.
\begin{figure}[top]
\begin{center}
\includegraphics[width=0.6\textwidth, ]{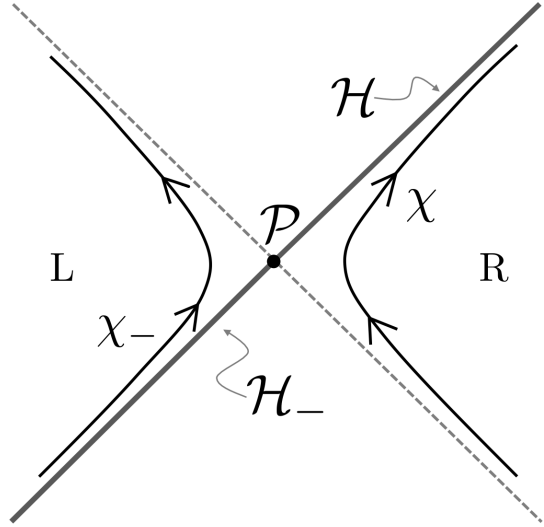}
\caption{
${\cal P}$ is an $(n-2)$-dimensional space-like hypersurface.
A null hypersurface ${\cal H}_-$ was considered as the LCH in \cite{Jacobson:1995ab}.
The region ${\rm L}$ in space-time behind the LCH is considered as 
a system with the Unruh temperature for an observer $\chi_-$ approaching asymptotically to ${\cal H}_-$.
}
\label{LCH}
\end{center}
\end{figure}
The local causal horizon is constructed as the past of these null vectors. Hence if we write the affine parameter of each null vector as $\lambda_K$ and the
coordinates of ${\cal P}$ as $y^A$ ($A=1, \cdots ,n-2$),
LCH ${\cal H}_-$ is a null hypersurface parametrized by $(y^A, \lambda_K)$.
Here we set $\lambda_K=0$ on ${\cal P}$.
The region $ {\rm L}$ in space-time behind the LCH is considered as a system whose temperature is identified with the Unruh temperature perceived by the uniformly accelerated observer $\chi_-$.
Jacobson derived the Einstein equation as the Clausius relation (\ref{Clausius}) where the entropy $S$ is proportional to the cross-sectional area of the null hypersurface $A=\int \sqrt{\gamma} d^{n-2}y$ while the heat is given by the flow of energy into the region $\rm L$ across the LCH $\delta Q/T=-2\pi \int_{{\cal H}_-} \lambda_K T_{\mu\nu} K^\mu K^\nu \sqrt{\gamma} d^{n-2}y d\lambda_K$.
$\gamma=\det (\gamma_{\mu\nu})$ is the determinant of the induced metric on the cross-section.
The normalization of the acceleration of an observer cannot be globally fixed, but
such an ambiguity is canceled with an ambiguity of measuring the heat $\delta Q$.
An important assumption in the derivation is the instantaneous equilibrium condition, namely the condition that the expansion $\theta$ and the shear $\sigma_{\mu\nu}$ should vanish on $\cal P$. This condition is necessary for the Clausius relation to be compatible with the Einstein equation at the lowest order of $ \lambda_K$.
Such an equilibrium condition is reasonable for a construction of the space-time thermodynamics.

The derivation was generalized to $F(R)$ gravity and Brans-Dicke theory by including  additional propagating degrees of freedom in $\delta Q$ distinguished from dissipative non-zero shear effect $\delta N$ \cite{Eling:2006aw,Chirco:2009dc,Chirco:2010sw}.
The method \cite{Eling:2006aw} was also applied to a scalar-tensor gravity in \cite{Cai:2006rs}.
But further generalizations to higher derivative theories 
are not yet successful,
suggesting that some refinement of the formulation will be necessary.\footnote{See \cite{Cai:2005ra,Akbar:2006er,Elizalde:2008pv,Brustein:2009hy,Bamba:2009gq,Piazza:2010hz,Yokokura:2011za} 
for other related works.}

Furthermore, there is a fundamental issue in the original derivation.
Jacobson considered an observer $\chi_-$ approaching ${\cal H}_-$ (the past half of the horizon) asymptotically from the region ${\rm L}$.
It is, however, more natural to consider an observer $\chi$ approaching  ${\cal H}$ (the future half of the horizon) asymptotically from the other region $ {\rm R}$ because the LCH is the boundary of causal regions for such an observer.
Under the null energy condition, an observer $\chi$ perceives positive energy flow across $\cal H$ {\it out of} the
region R while an observer $\chi_{-}$ perceives positive energy flow across  $\cal H_{-}$ {\it into} the region L.
Namely the sign of  the heat flow $\delta Q$ discussed in the previous paragraph 
is opposite  to each other because of  an opposite sign of  $\lambda_K$.
On the other hand, in either case, the area change  is given by $\delta(A/4) = \frac{1}{4} \int \theta \  d^{n-2}y d\lambda_K = \frac{1}{4} \int \{ -R_{\mu\nu}K^{\mu}K^{\nu}|_{\cal P}\ \lambda_K +{\cal O}(\lambda_K ^2) \}d^{n-2}y d\lambda_K$ under the equilibrium condition.
Thus one would obtain the Einstein equation with a wrong sign of the Newton constant 
if the observer $\chi$ were used in the original derivation instead of $\chi_-$.

In order to solve the issue of the choice of an observer, 
Parikh and Sarkar \cite{Parikh:2009qs} made use of the Noether charge method and introduced an entropy in an observer dependent way. 
It was further refined  in \cite{Guedens:2011dy}.
Padmanabhan \cite{Padmanabhan:2009ry} also suggested a method using the Noether current and discussed 
how one can justify such  derivations.
They succeeded to generalize the derivation to theories whose Lagrangians are made from Riemann tensors but without their derivatives.
But the entropy used in the derivation depends on the approximate Killing vectors they introduced, 
and  its relation to the original derivation is not clear. 

In this Letter,
we consider a null hypersurface with  a non-vanishing expansion and  a shear  on ${\cal P}$
and give a modification of the original derivation of the Einstein equation.
Such a generalization turns out to be inevitable since these quantities evolve and cannot be set zero  
after leaving the hypersurface ${\cal P}$ along $\lambda_K$ if there is a heat flow across ${\cal H}$. 
We show that this gives an important hint to solve the fundamental issue in the original derivation.

In section 2, we introduce a new definition of the entropy change in the Clausius relation on a general null hypersurface 
with a non-vanishing $\theta$,
and give an alternative derivation of the Einstein equation of state.
In section 3, we interpret the derivation based on the Noether charge method and explain why it works.
It also clarifies what was missing in the original formulation by Jacobson.
In section 4, we consider a generalization to $F(R)$ gravity.
Section 5 is devoted to conclusions and discussions.
We comment on yet another derivation of the Einstein equation of state.
We also discuss a possibility  to formulate the space-time thermodynamics in an observer dependent way.

\section{The Einstein equation of state}

\hspace{16pt}We introduce a modified version of the derivation of the Einstein equation of state from a thermodynamic relation.
Like the original derivation by Jacobson, we first introduce a
 space-like hypersurface ${\cal P}$ and future directed null vectors $k^\mu$ perpendicular to the surface ${\cal P}$. 
However, instead of  an observer $\chi_-$ and the local horizon ${\cal H}_-$ in Figure \ref{LCH},
we consider ($\chi$,${\cal H}$) and compare the change of area of ${\cal P}$ with the heat flow across ${\cal H}$. 
We do not either impose the instantaneous equilibrium condition $\theta=\sigma_{\mu\nu}=0$ below.

Null vectors are parametrized by the affine parameter $\lambda_K$,
$K^\mu = (\partial/\partial \lambda_K)^\mu$. The entropy change was defined in \cite{Jacobson:1995ab}
as a change of area $\delta S \propto \partial \sqrt{\gamma}/\partial \lambda_K $ with respect to $\lambda_K$.
We will use a different parameter $\lambda$ related to $\lambda_K$ by
\begin{align}
\frac{\partial \lambda_K}{\partial \lambda}= e^{c}\ , \notag
\end{align}
where $c$ is some function which will be determined below.
The null vector along the direction of $\lambda$ is written as $k^\mu =(\partial /\partial \lambda)^{\mu} =e^{c} K^\mu$ and  satisfies $k \cdot \nabla k^\mu =c' k^\mu$.
Here prime  stands for a derivative with respect to $\lambda$ .

Before discussing the thermodynamic relation in the presence of an energy flow, 
we will first derive a kinematical\footnote{We use the word ``kinematical" when we talk about the effect caused by a non-zero value of the expansion at the hypersurface ${\cal P}$. 
On the contrary, in the presence of an energy flux, 
the curvature is expected to become non-zero and the evolution of the expansion is determined by the Raychaudhuri equation with the Ricci tensor.
Hence it becomes dynamical.} thermodynamic relation in flat space without an energy flow.
Since the expansion of the area is not assumed to vanish, i.e.
\begin{align}
\theta = (k \cdot \partial) \ln \sqrt{\gamma} =\frac{d \ln \sqrt{\gamma}}{d \lambda} \neq 0\ , \notag
\end{align}
we need to compensate such a kinematical 
change of the area by a local scale transformation 
of the parameter of the observer's world line.
It changes the definition of the acceleration from the ordinary one measured with respect to the affine parameter $\lambda_K$ to another measured by $\lambda$.
In the next section, 
we show that such a change of acceleration, and accordingly temperature, corresponds to taking a different observer approaching asymptotically to the null hypersurface ${\cal H}$.
We thus introduce the following quantity as a change of ``entropy density"\footnote{The reader may suspect why such a rescaling of ``entropy" density is necessary.  It comes from the fact that the quantity we are considering in a thermodynamic relation is a combination of the temperature $T$ and the entropy $\delta S$.
In a diffeomorphism invariant theory of gravity, we can always change the local scaling of time of an observer and accordingly we need to consider the effect on temperature.
In this expression, we absorbed the effect of such a local scaling of observer's time into the definition of the ``entropy" density. It is finally justified in (\ref{LchiNcharge}) where we derive the thermodynamic relation in this section from the most general identity of the Noether charge.}
\begin{align}
 T \delta S \equiv \frac{\kappa}{2 \pi} (k \cdot \partial)
\left[ e^{-c} \frac{\sqrt{\gamma}}{4}\right]
d^{n-2}y d\lambda \ .
\label{entropychange}
\end{align}
where $c=\lambda \theta(\lambda,y)+{\cal O}(\lambda^3)$ and $\kappa$ is a constant which is  related, 
but not identical,
 to the acceleration of an observer approaching the null hypersurface ${\cal H}$ asymptotically.
The above $T \delta S$ is proportional to $(\theta - c')$ and  vanishes at the leading order in an expansion of $\lambda$.
Hence (\ref{entropychange}) is considered to be a natural generalization of the "entropy" change  for $\theta \neq 0$ null hypersurfaces.\footnote{If the hypersurface has an expansion, we need to rescale the acceleration of the
corresponding observer so that the product of the area and the acceleration becomes constant. It is why
we need an extra factor $e^{-c}$ in $T\delta S.$ 
}
But as we will see below, this makes a big change at  higher orders in the expansion of $\lambda$.
We will see in the next section that such a special choice of $c$ corresponds to considering a specific accelerating observer that asymptotes to ${\cal H}$.
We also give an interpretation of  (\ref{entropychange}) based on the Noether charge method.

Let us evaluate higher order terms of  $\lambda$ when there is  no energy flux in the flat space-time.
The Raychaudhuri equation is given by
\begin{align}
\theta' = c' \theta -\frac{\theta^2}{n-2} - \sigma_{\mu\nu} \sigma^{\mu\nu} \ .
\label{Raych}
\end{align}
Then (\ref{entropychange}) becomes
\begin{align}
T \delta S = - \frac{1}{8 \pi}
\theta' \sqrt{\gamma} d^{n-2}y \ (\kappa \lambda) d\lambda + {\cal O}(\lambda^2)
= \delta N + {\cal O}(\lambda^2) \notag
\end{align}
where
\begin{align}
\delta N =  \frac{1}{8\pi} \left( -\frac{n-3}{n-2} \theta^2 +\sigma^2 \right) \sqrt{\gamma} d^{n-2}y \ (\kappa \lambda )d\lambda \notag.
\end{align}
It is the same as the entropy production term in the membrane paradigm \cite{Thorne}.
That the correct viscous coefficients are reproduced makes the definition  (\ref{entropychange})
plausible. 
 
If we imposed an instantaneous equilibrium condition with $\theta=\sigma=0$ on ${\cal P}$,  
it seems that it was sufficient to consider the change of area itself as  $T\delta S$.
But that would lead to an opposite sign of the curvature term as shown below.\footnote{Speaking more rigorously, the reason why the original derivation of Jacobson could not 
give the correct sign for the Einstein equation is the following. 
We can always consider a hypersurface with a vanishing expansion at a point $p$. 
However, it cannot be set zero away from the point $p$ with $\lambda \neq 0$.
In the original derivation, only a half of the effects of the evolution of $\theta(\lambda)$ was taken into account.
The other half, namely, the effect of the local rescaling of temperature associated with a non-vanishing $\partial \theta(\lambda=0)$ was not considered. Because of this, the coefficient of the curvature term became opposite.  
This is the reason why he needed to consider an observer in the left wedge instead of that in the right wedge. 
 }

We then require the Clausius relation to hold  in the presence of an energy flux;
\begin{align}
T \delta S -\delta N = \delta Q \label{Clausius2}
\end{align}
where the energy flux is given by
\begin{align}
\delta Q = T_{\mu\nu} k^\mu k^\nu \sqrt{\gamma} d^{n-2}y \ (\kappa \lambda 
)d\lambda \ . \notag
\end{align}
In this case, the curvature term $R_{\mu\nu} k^{\mu} k^{\nu}$ must be included in the Raychaudhuri equation 
(\ref{Raych})
and the Clausius relation (\ref{Clausius2}) gives a relation 
\begin{align}
 R_{\mu\nu} k^\mu k^\nu = 8 \pi  T_{\mu\nu} k^\mu k^\nu \ . \notag
\end{align}
Since the direction of the null vector $k^\mu$ is arbitrary at point $p$
and the energy momentum tensor satisfies $\nabla_\mu T^{\mu\nu}=0$},
we can obtain the Einstein equation 
\begin{align}
 R_{\mu\nu} - \frac{R g_{\mu\nu}}{2} =8\pi T_{\mu\nu} - \Lambda g_{\mu\nu} \ . \notag
\end{align}
where the value of the cosmological constant $\Lambda$ can be chosen freely.

In the above discussions, we changed the definition of $T \delta S$ by selecting a special  observer.
Accordingly $T\delta S$ becomes proportional to 
$\theta -c'=-\lambda (\partial \theta/\partial \lambda)$.
This minus sign gives the correct sign of $\delta N$ and $\delta Q$ for 
the observer $\chi$ that asymptotes to ${\cal H}$ from $ {\rm R}$ in Figure \ref{LCH}.
On the contrary, as in the original derivation,
if we expand the area in $\lambda_K$ with a condition $\theta_{K}=\partial \ln \sqrt{\gamma}/\partial \lambda_K=0$ at $\cal P$,
it gives a term $\lambda_{K} (\partial \theta_{K}/\partial \lambda_{K})$.
Note that the sign in front is opposite, which gives an opposite sign of the curvature term.
This is the reason why it was necessary to consider an observer $\chi_-$ that asymptotes to
 ${\cal H}_-$ .
The extra factor $e^{-c}$ played two important roles. The first is to compensate the 
expansion of the area at ${\cal P}$. 
It is the leading order effect with respect to $\lambda$. 
The next role is the next-to-leading order effect.
It makes the sign of the term $-\lambda \theta'$ opposite.
If there is an  energy flow, the expansion of the area no longer vanishes
away from ${\cal P}$ (the next-to-leading order effect of $\lambda$)
 and we cannot neglect the effect of the expansion. 
It is the reason why an extra term $e^{-c}$ is inevitable to derive the Einstein equation  from the Clausius relation.

The derivation of the Einstein equation in this section looks ad hoc, 
but suggests that the space-time thermodynamics inevitably becomes  observer dependent.
We give an interpretation of the above derivation and the observer dependent definition of $T\delta S$ based on the Noether charge method in the next section.

\section{Interpretation based on the Noether charge method}

\subsection{Noether charge}

\hspace{16pt}The Noether charge method \cite{Wald:1993nt} gives a fundamental relation between the Noether charge
and thermodynamic quantities for  general diffeomorphism invariant theories of gravity.
Especially it defines the Wald entropy of a black hole and 
leads to the first law of thermodynamics. 
Its applications so far have been mostly limited to black holes with the Killing horizons,
but the method can play an important role in constructing the space-time thermodynamics beyond black holes.

We first review the construction of the Noether charge in this subsection.
Given a diffeomorphism invariant theory of gravity with the  Lagrangian $n$-form $\bf L$,
its variation under $\delta g^{\mu\nu}$ is given by
\begin{align}
\delta {\bf L} =  {\boldsymbol \epsilon} \frac{{\cal G}_{\mu\nu}}{16\pi} \delta g^{\mu\nu} + {\bf d}{\bf \Theta}(g, \delta g) \ . \notag
\end{align}
where ${\boldsymbol \epsilon}$ is a volume $n$-form.
${\cal G}_{\mu\nu}$ is a generalization of the Einstein tensor satisfying $\nabla_\mu {\cal G}^{\mu\nu}=0$.
Combined with the variation of matter field, the equation of motion becomes ${\cal G}_{\mu\nu} = 8 \pi T_{\mu\nu}$.
For a general coordinate transformation $\delta g^{\mu\nu}= {\cal L}_\chi g^{\mu\nu}$ generated by a vector field $\chi$,
diffeomorphism invariance of the theory gives $\delta {\bf L}={\bf d}i_\chi {\bf L}$ and there exists a Noether charge $(n-2)$-form ${\bf Q}_\chi$ such that the Noether current $(n-1)$-form ${\bf J}_{\chi}$ can be written as
\begin{align}
{\bf J}_{\chi} \equiv
{\bf \Theta}(g,{\cal L}_{\chi}g)- i_{\chi}{\bf  L}={\bf dQ}_{\chi}+\frac{1}{8\pi}{\cal G}^{\mu\nu}\chi_{\mu}{\boldsymbol \epsilon}_{\nu} \ . \label{Noethercurrent}
\end{align}
where ${\boldsymbol \epsilon}_\nu$ is an $(n-1)$-volume form.
The construction of the Noether charge is quite general but the definition of ${\bf Q}_\chi$ depends explicitly on the choice of the vector $\chi$. The choice corresponds to a choice of an observer who observes the energy flux and measures the change of area.
In the case of black holes,
we can take the Killing vector as $\chi$.
This corresponds to measuring the energy flux by an observer sitting at $r=\infty$.
In deriving the Einstein equation of state in general space-time, we need to choose an appropriate observer so that the thermodynamic relation becomes as simple as possible.

\subsection{The choice of observers}

\begin{figure}[top]
\begin{center}
\includegraphics[width=0.6\textwidth, ]{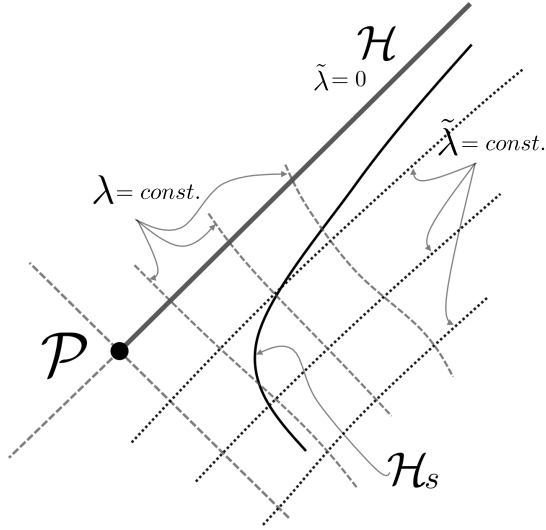}
\caption{
${\cal P}$ is an $(n-2)$-dimensional space-like hypersurface containing the point $p$.
The null hypersurface ${\cal H}$ is defined by $\tilde{\lambda}(x)=0$,
${\cal H}_s$ is defined by $\lambda \tilde{\lambda}=-s^2/2$.
}
\label{patch}
\end{center}
\end{figure}

\hspace{16pt}We then  define the tangent vector $\chi^\mu$ of an observer by generalizing the uniformly accelerated observer in flat space-time. First we define a function $\tilde{\lambda}(x)$ such that each of $\tilde{\lambda}(x)=const.$ hypersurface is a null hypersurface. 
Especially $\tilde{\lambda}(x)=0$ is set as the null hypersurface ${\cal H}$ containing the point $p$ in Figure \ref{patch}.
Then its  normal vector $K^\mu = -g^{\mu\nu}\partial_\nu \tilde{\lambda}$ is shown to satisfy the equation $(K \cdot \nabla) K^\mu =0$.
Thus $K^\mu$ is affine tangent to the null hypersurface ${\cal H}$ and we write the null vector by an affine parameter $\lambda_K$ as
\begin{align}
K^\mu = \left( \frac{\partial}{\partial \lambda_K} \right)^\mu. \notag
\end{align}
This defines a function $\lambda_K(x)$ in the space-time.
We can choose the hypersurface $\lambda_K (x)=0$ to be null and contain the point $p$.
The space-like surface with a fixed $\tilde{\lambda}$ and $\lambda_K$ is parametrized by $y^A , A=1,\cdots , n-2$.
The space-like surface $\tilde{\lambda}=\lambda_{K}=0$ is identified as $\cal P$ and $p$ is the point $y^{A}=0$ in this surface.
Note that other hypersurfaces $\lambda_K (x)= const. \neq 0$ are not always null. 
 
In the flat space-time with a metric $ ds^2=-2d\lambda_K d\tilde{\lambda}$ on the $(\lambda_K, \tilde{\lambda})$ plane,
an observer whose world line is given by $\lambda_K \tilde{\lambda}=-s^2/2=const.$ 
and $y^A=const.$ represents a uniformly accelerated observer,
and $s \rightarrow 0$ limit asymptotes to the null hypersurface ${\cal H}$.
We define another type of observers who asymptotes to the null hypersurface ${\cal H}$ by using a new coordinate  $\lambda$.
The new coordinate  $\lambda$ is defined to be related to $\lambda_K$ as
\begin{align}
\frac{\partial \lambda_K}{\partial \lambda}=e^{c} \notag
\end{align}
with  a condition $c(\lambda=0)=0$.
Then we define  world lines of a set of new observers by the relations $y^A=const.$ and
\begin{align}
\lambda \tilde{\lambda}=-\frac{1}{2}s^2 \ . \label{observer}
\end{align}
The condition (\ref{observer}) defines an  $(n-1)$-dimensional hypersurface ${\cal H}_s$.
The limit $s \rightarrow 0$ of ${\cal H}_s$ asymptotes to the null hypersurface ${\cal H}$.
An $(n-2)$-dimensional space-like hypersurface ${\cal S}_{(\lambda, \tilde{\lambda})}$ with a fixed $\lambda$ and $\tilde{\lambda}$ 
is parametrized by $y^A$.
We define  a null vector
\begin{align}
k^\mu  \equiv \left( \frac{\partial }{\partial \lambda} \right)^\mu_{(\tilde{\lambda},y^A)} = e^{c} K^\mu \notag
\end{align}
and space-like vectors
\begin{align}
e_{A}^{\mu} \equiv \left( \frac{\partial }{\partial y^A} \right)^\mu_{(\lambda, \tilde{\lambda})} . \notag
\end{align}
We also introduce another null vector $l^\mu$ which satisfies
\begin{align}
l\cdot k=-1\ ,\ l\cdot e_{A}=0\ ,\  l\cdot l=0 \ . \notag
\end{align}
Note that it is generally different from $\partial/\partial \tilde{\lambda}$.
The induced metric $\gamma_{\mu\nu}$ on ${\cal S}_{(\lambda, \tilde{\lambda})}$ is given by $\gamma_{\mu\nu}=g_{\mu\nu}+k_\mu l_\nu +k_\nu l_\mu$.

The normal vector $\beta$ to the hypersurface ${\cal H}_s$ is given by 
\begin{align}
\beta_{\mu}&=-\kappa \partial_{\mu}(\lambda \tilde{\lambda}) \notag \\
&=\kappa l\cdot \partial (\lambda \tilde{\lambda})k_{\mu}+\kappa k\cdot \partial (\lambda \tilde{\lambda})l_{\mu} \notag \\
&=\kappa (e^{-c}\lambda +\tilde{\lambda} l\cdot \partial \lambda )k_{\mu}+\kappa \tilde{\lambda}l_{\mu} \notag \\
&\xrightarrow{\tilde{\lambda}\to 0} \kappa e^{-c} \lambda k_{\mu} \ , \notag
\end{align}
where $\kappa$ is an arbitrary positive constant which has the dimension of $(\text{length})^{-1}$.
Here we have used $(l \cdot \partial)\tilde{\lambda} = -l\cdot K =e^{-c}$.
The last line gives a  limiting form of $\beta_\mu$ on $\cal H$.

The tangent vector $\chi$ of ${\cal H}_s$ is determined to be orthogonal to $\beta$ and $e_A$.
If we choose its normalization so that it coincides with $\beta$ on $\cal H$,
we have
\begin{align}
\chi^{\mu}&=\kappa (e^{-c}\lambda +\tilde{\lambda} l\cdot \partial \lambda )k^{\mu}-\kappa \tilde{\lambda}l^{\mu} 
\label{chi} \\
&\xrightarrow{\tilde{\lambda}\to 0} \kappa e^{-c} \lambda k^{\mu} \ . \notag
\end{align}

\subsection{Thermodynamic relation}

\hspace{16pt}We now derive a thermodynamic relation for the observer $\chi$ from the Noether charge method.
In order to compare with the result in the  previous section,
we  consider the Einstein-Hilbert Lagrangian for simplicity.
By operating $i_{\chi}$ on the Noether current (\ref{Noethercurrent}) and using the formula $i_{\chi}{\bf dQ}_{\chi}={\cal L}_\chi {\bf Q}-{\bf d}i_\chi {\bf Q}$,
we obtain the following relation 
\begin{align}
{\cal L}_{\chi}{\bf Q}_{\chi} -{\bf d}i_\chi {\bf Q} -i_{\chi} {\bf \Theta}(g,{\cal L}_{\chi}g)
&=\frac{1}{8\pi}\chi^{\mu}{\cal G}^{\nu}_{\mu}\chi^{\rho}{\boldsymbol \epsilon}_{\rho\nu} \ . \label{pre_Clausius}
\end{align}
This is the most general identity that holds for any observer $\chi$.
In the following we choose a special observer $\chi$ so that the 
identity becomes simpler.
Each term shall be evaluated on an $(n-2)$-dimensional space-like surface 
${\cal S}_{(\lambda,\tilde{\lambda})}$, 
and ${\boldsymbol \epsilon}_{\mu\nu}$ in the RHS of (\ref{pre_Clausius}) is the binormal to ${\cal S}_{(\lambda, \tilde{\lambda})}$,
\begin{align}
{\boldsymbol \epsilon}_{\mu\nu} =2 k_{[\mu}l_{\nu]} \sqrt{\gamma} \ {\bf d}y^1 \wedge \cdots \wedge {\bf d}y^{n-2}. \notag
\end{align}
For the Einstein-Hilbert Lagrangian, the Noether charge and the surface term are given by
\begin{align}
{\bf Q}_\chi & =\frac{-1}{16 \pi}\nabla^{[\mu} \chi^{\nu]} {\boldsymbol \epsilon}_{\mu\nu} \ , \label{Ncharge} \\
{\bf \Theta}(g, \delta g) &=\frac{1}{8\pi} g^{\nu [\rho} \nabla^{\mu]} \delta g_{\mu\nu} {\boldsymbol \epsilon}_\rho \ . \notag
\end{align}
For the observer $\chi$ defined in (\ref{chi}),
the Noether charge (\ref{Ncharge}) on ${\cal S}_{(\lambda,\tilde{\lambda})}$ becomes 
\begin{align}
{\bf Q}_\chi \xrightarrow{\tilde{\lambda} \rightarrow 0} \frac{\kappa e^{-c}}{8 \pi} \sqrt{\gamma} d^{n-2}y \ . \label{Ncharge2}
\end{align}
Then the first term of (\ref{pre_Clausius}) is written as
\begin{align}
{\cal L}_{\chi}{\bf Q}_\chi \xrightarrow{\tilde{\lambda} \rightarrow 0} &
\frac{\kappa^2 }{8 \pi} e^{-2c}\lambda \left( \theta-k\cdot \partial c \right)\sqrt{\gamma} d^{n-2}y \label{LchiNcharge} \\
=&\frac{\kappa }{2 \pi} \ \chi \cdot \partial \left( e^{-c}\frac{\sqrt{\gamma}}{4} \right) d^{n-2}y \notag
\end{align}
where $\theta$ is the expansion of the null vector $k^\mu$ on ${\cal H}$.
Hence, by multiplying $dt$, (\ref{LchiNcharge}) becomes $T\delta S$ defined in (\ref{entropychange}).
Here $t$ is the time variable generating the tangent vector $\chi =\partial/\partial t$,
and $dt=(\kappa \lambda e^{-c})^{-1} d \lambda$.
The second term ${\bf d}i_\chi {\bf Q}_\chi$ in (\ref{pre_Clausius}) vanishes in the limit of $\tilde{\lambda} \rightarrow 0$.
The third term is more complicated.
In the derivation of the Wald formula of black hole entropy, this term vanishes on the bifurcation surface since the Killing vector vanishes there.
In our setting, however, the space-time does not generally have a Killing vector and we cannot drop the term.
Since we are considering a null hypersurface with a non-vanishing expansion,
the entropy production term $\delta N$ is expected to appear from this term.
Indeed a straightforward calculation shows
\begin{align}
i_{\chi} {\bf \Theta}(g,{\cal L}_{\chi}g)
\xrightarrow{\tilde{\lambda} \to 0}
 \frac{\kappa^2}{8\pi} e^{-2c} \lambda \bigg[ &
\theta-k\cdot \partial c \notag \\
&+\lambda\left( k\cdot \partial \theta +\frac{1}{n-2}\theta^2+\sigma^2 -\theta k\cdot \partial c\right) \bigg] \sqrt{\gamma}d^{n-2}y. \notag
\end{align}
If we take a special choice $c \xrightarrow{\tilde{\lambda} \to 0} \lambda \theta(\lambda,y)+{\cal O}(\lambda^3)$,
it is simplified as
\begin{align}
i_{\chi} {\bf \Theta}(g,{\cal L}_{\chi}g)|
 \xrightarrow{\tilde{\lambda} \rightarrow 0} &\frac{(\kappa\lambda e^{-c})^2}{8\pi} \left( -\frac{n-3}{n-2}\theta^2+\sigma^2 \right) \sqrt{\gamma}d^{n-2}y  +{\cal O}(\lambda^3) \notag \\
&=\frac{1}{8\pi} \left( -\frac{n-3}{n-2}\hat{\theta}^2+\hat{\sigma}^2 
\right) \sqrt{\gamma}d^{n-2}y  +{\cal O}(\lambda^3) \label{Thetaterm}
\end{align}
where $\hat{\theta}$ and $\hat{\sigma}_{\mu\nu}$ are respectively the expansion and shear of $\chi$ on ${\cal H}$.
This is nothing but the entropy production term 
of $(n-2)$-dimensional fluid with a bulk viscosity $-(n-3)/8\pi(n-2)$ and a shear viscosity $\eta=1/16\pi$,
and consistent with the values obtained in the membrane paradigm picture of black holes.
The term that vanished in the term $i_\chi {\bf \Theta}$ by taking the special choice of the observer $\chi$ may be interpretable as the change of ``temperature".\footnote{The ``temperature"  should be a rescaled temperature
which is proportional, not to the acceleration of the observer itself, but to
the acceleration  multiplied by a factor $e^{\lambda \theta}$. 
Such a rescaling is necessary 
 to compensate the increase of the area at $\lambda=0$ on ${\cal P}$.
Since the Noether charge on a cross-section of ${\cal H}$ gives a combination of $TS$
and cannot be dissociated at the classical level, 
there may exist an arbitrariness  in  interpreting  the Noether charge relation as a thermodynamic one.}
This is indeed true in the case of the derivation of the first law in the black hole thermodynamics \cite{Bardeen:1973gs}.

Collecting the above contributions of (\ref{LchiNcharge}) and (\ref{Thetaterm}),
and multiplying the relation by $dt$,
the Noether charge relation (\ref{pre_Clausius}) becomes
\begin{align}
T \delta S - \delta N &= \frac{1}{8\pi}\chi^{\mu}{\cal G}_{\mu\nu}k^{\nu}\sqrt{\gamma} d^{n-2}y d\lambda \notag
\end{align}
where $T\delta S$ is defined in (\ref{entropychange}).

If we view
 the original derivation by Jacobson \cite{Jacobson:1995ab} in the Noether charge method,
the contribution of the term $i_{\chi} {\bf \Theta}(g,{\cal L}_{\chi}g)$ is neglected.
But as we saw, both terms, ${\cal L}_{\chi}{\bf Q}_{\chi}$ and $i_{\chi} {\bf \Theta}(g, {\cal L}_{\chi}g)$,
contain a term proportional to  $R_{\mu\nu}k^\mu k^\nu$.
If we set $c=0$, 
the coefficient of the curvature term is (-1) in the former and (+2) in the latter.
Then if we did not include the term $i_{\chi} {\bf \Theta}(g, {\cal L}_{\chi}g)$,
we would have a negative sign for the energy flow.
This is the reason why it was necessary to take the observer $\chi_-$ in the original derivation.
The inclusion of the term $i_\chi {\bf \Theta}$ is necessary  for the correct sign of the energy flux but also for the correct coefficient of the entropy production term $\delta N$.
Instead, if we set $c=\lambda \theta$,
the term $i_{\chi} {\bf \Theta}(g, {\cal L}_{\chi}g)$ is reduced to
$\delta N$ and we can get the correct coefficient from ${\cal L}_{\chi}{\bf Q}_{\chi}$.
Hence we can take the natural observer $\chi$, instead of $\chi_-$,
to generate the  Einstein equation.

\section{$F(R)$ gravity}

\hspace{16pt}The method shown in the previous sections can be extended to higher derivative theories of gravity,
but a subtlety arises in a choice of  the  entropy production term $\delta N$.
In this section we focus  on the $F(R)$ theory of gravity.

Eq. (\ref{pre_Clausius}) holds in general,
but in the case of $F(R)$ gravity the Noether charge ${\bf Q}$ has two terms
\begin{align}
{\bf Q}_\chi={\bf X}^{\mu\nu}\nabla_{[\mu}\chi_{\nu]}+{\bf W}^\mu\chi_\mu
= \frac{1}{16 \pi} \left( -f(R) \nabla^{[\mu} \chi^{\nu]} + 2 \chi^{[\nu} \nabla^{\mu]} f(R) \right) {\boldsymbol \epsilon}_{\mu\nu} \notag
\end{align}
where $f(R)=\partial F(R)/ \partial R$.
The Lagrangian is given by $F(R)/(16\pi)$.
On the bifurcation surface of a black hole, the second term vanishes and the Wald entropy is given by only the first term.
Here we also regard the first term as a contribution to  $T \delta S$
and the second as a part of ${\delta N}$.
The term ${\bf d} i_\chi {\bf Q}$  vanishes\footnote{In more general theories, the term does not vanish 
and  contribute to the thermodynamic relation as an $(n-3)$-dimensional surface effect.} when we take $\tilde{\lambda} \rightarrow 0$ and Eq. (\ref{pre_Clausius}) becomes
\begin{align}
{\cal L}_{\chi}({\bf X}^{\mu\nu}\nabla_{[\mu}\chi_{\nu]} )-\left[ i_{\chi} {\bf \Theta}(g,{\cal L}_{\chi}g) -{\cal L}_{\chi}({\bf W}^\mu \chi_\mu)\right]
=\frac{1}{8\pi}\chi^{\mu}{\cal G}^{\nu}_{\mu}\chi^{\rho}{\boldsymbol \epsilon}_{\rho\nu} \ . \label{F(R)}
\end{align} 
For the observer with the tangent vector $\chi$,
the first term of (\ref{F(R)}) becomes
\begin{align}
{\cal L}_{\chi}({\bf X}^{\mu\nu}\nabla_{[\mu}\chi_{\nu]} )\xrightarrow{\tilde{\lambda} \rightarrow 0} & \frac{\kappa^2}{8\pi} e^{-2c} \lambda f \left[ \tilde{\theta}-k\cdot \partial c\right] \sqrt{\gamma}d^{n-2}y \label{F(R)LchiX}\\
=& \frac{\kappa}{2\pi} \ \chi \cdot \partial \left[ e^{-c}f \frac{\sqrt{\gamma}}{4} \right] d^{n-2}y \ ,\notag
\end{align}
where $\tilde{\theta}=\theta + k\cdot \partial \ln f =\theta+f'/f$. 

The combination of two terms in the square bracket in (\ref{F(R)}) becomes
\begin{align}
i_{\chi} {\bf \Theta}& (g,{\cal L}_{\chi}g) -{\cal L}_{\chi}({\bf W}^\mu \chi_\mu) \notag  \\ 
 &\xrightarrow{\tilde{\lambda} \rightarrow 0} \frac{\kappa^2}{8\pi} e^{-2c} \lambda f \bigg[ \tilde{\theta}-k\cdot \partial c 
+\lambda\Big{\{ } k\cdot \partial \tilde{\theta}+\frac{1}{n-2}\tilde{\theta}^2+\tilde{\sigma}^2 \label{F(R)deltaN}\\
&\hspace{50pt}+\frac{n-1}{n-2}(k\cdot \partial \ln f)^2 -\tilde{\theta}\Bigl( k\cdot \partial c+\frac{2}{n-2}k\cdot \partial \ln f \Bigr) \Big{\} } \bigg]\sqrt{\gamma}d^{n-2}y \ . \notag 
\end{align}
There is an arbitrariness in the choice of  $c$,
so we determine it to make (\ref{F(R)deltaN}) as simple as possible.
We also demand that the lowest order term in (\ref{F(R)LchiX}) vanishes as before.
Then $c$ can be chosen as
\begin{align}
c &\xrightarrow{\tilde{\lambda} \rightarrow 0}\lambda \tilde{\theta}_{| \lambda=0}+\lambda^2 \left( k\cdot \partial \tilde{\theta} -\frac{1}{n-2}\tilde{\theta} k\cdot \partial \ln f\right)_{| \lambda=0} +{{\cal O}(\lambda^3)}\ . \label{Observer_F(R)}
\end{align}
Plugging it into (\ref{F(R)deltaN}), we obtain
\begin{align}
 i_{\chi} {\bf \Theta}& (g,{\cal L}_{\chi}g) -{\cal L}_{\chi}({\bf W}^\mu \chi_\mu)  \label{F(R)EP} \\
&\xrightarrow{\tilde{\lambda} \rightarrow 0} \frac{(\kappa\lambda)^2}{8\pi} f \left( -\frac{n-3}{n-2}\tilde{\theta}^2 + \sigma^2+\frac{n-1}{n-2}(k\cdot \partial \ln f)^2 \right) \sqrt{\gamma}d^{n-2}y+{\cal O}(\lambda^3). \notag
\end{align}
The third term is nothing but the contribution from the additional propagating degrees of freedom which is treated as an extra heat in the previous derivations
\cite{Eling:2006aw,Chirco:2009dc,Chirco:2010sw}.

Based on the above arguments, we can derive the equation of motion for $F(R)$ gravity
from the thermodynamic relation:
Define the entropy change $T\delta S$ and the entropy production $\delta N$ as (\ref{F(R)LchiX}) and (\ref{F(R)EP}) respectively and require the Clausius relation $T\delta S -\delta N =\delta Q$ to hold for an observer approaching asymptotically to the null hypersurface ${\cal H}$.
The observer is taken to have a tangent vector $\chi$ with $c$ defined in (\ref{Observer_F(R)}).
Then the Clausius relation gives the equation of motion for $F(R)$ gravity by using the Raychaudhuri equation,
the second Bianchi identity and the relation $\nabla^\mu\nabla_\nu \nabla_\mu f=\partial_\nu \Box f +R_{\nu\mu}\partial^\mu f$.

\section{Conclusions and discussions}

\hspace{16pt}In this Letter we proposed an alternative derivation of the Einstein equation of state starting from a modified Clausius relation. 
The hypersurface is allowed to have a non-vanishing expansion and shear.
We therefore add an entropy production term in the Clausius relation: $T\delta S -\delta N =\delta Q$.
In order to be consistent with the  Raychaudhuri equation in flat space-time, we need to choose a special observer with a tangent vector field $\chi$.
Once we make the thermodynamic relation to hold in flat space-time, 
we could show that the Clausius relation generates the Einstein equation. 
It can be generalized to higher derivative gravities, and as an example we studied $F(R)$ gravity. 

We  also clarified the reason why it was necessary to consider an observer behind the local causal horizon in the original formulation by Jacobson.
In the Noether charge identity, the term $i_\chi {\bf \Theta}$ cannot be  generally neglected, except for a special case like the stationary black holes which have the Killing vector.
Since the term 
 contains a contribution of the curvature term $R_{\mu\nu} k^\mu k^\nu$,
we would get a wrong coefficient if we naively impose the Clausius relation $T\delta S = \delta Q$.
In order to cancel the contribution from the term $i_{\chi} {\bf \Theta}(g, {\cal L}_{\chi}g)$,
we need to take a special vector $\chi$.
This is the reason why it was necessary to modify the entropy change  $T\delta S$ by adding an effect of the observer dependent factor $e^{-c}$.

The derivation of the equation of motion needs a special choice of an
observer and it gives a factor $e^{-c}$ in the definition of $T \delta S$.
The factor is naturally understood from the Noether charge method, but we do not know
{\it a priori} which observer we should choose in constructing a thermodynamic relation in space-time.
The condition that the lowest order term in $T\delta S$  must vanish determines a partial 
form of $c$, but the rest depends on how we divide 
thermodynamic quantities into $T\delta S$ and $\delta N$. 
Another remaining issue is the definition of entropy in general theories of gravity.
In the case of $F(R)$ gravity, we considered the term proportional to $\nabla \chi$ 
as the entropy. It is consistent with the Wald formula of black hole entropies, but
it is not obvious if it is generally so.

In a recent paper \cite{Guedens:2011dy},
the importance of taking a special choice of observers is emphasized.
They introduced an approximate Killing vector field $\xi$ 
so that the term $i_{\xi} {\bf \Theta}(g, {\cal L}_{\xi}g)$ vanishes in the leading orders in $\lambda$.
They succeeded to derive the Einstein equation from the Clausius relation $T\delta S =\delta Q$
in a theory constructed from the metric and Riemann tensor
by considering an observer with the tangent vector $\xi$.
In order to extend the derivation to more general theories, one may need to extend the notion of 
entropy beyond the  Wald entropy.
An advantage of their  approach is that the choice of an observer is independent of the  gravity theories,
but one is required to consider a narrow region of the null hypersurface in which the approximate Killing
equation is satisfied. Furthermore, the observer's four velocity is not directly related with the 
null generator of the hypersurface.

Let us comment on the relation of the approach \cite{Guedens:2011dy} and ours.
The basic identity underlying the derivation of the Einstein equation in both approaches is the Noether charge identity (\ref{pre_Clausius}) or (\ref{Noethercurrent}).
This identity is very general and holds for any generally covariant theory of gravity. The issue of ``deriving the Einstein equation" is how we can interpret the Noether charge identity as a thermodynamic relation. 
The identity depends on the explicit choice of the four velocity vector $\chi$ of an observer. In this sense, it can have various different thermodynamic interpretations. The difference between our approach and the approach in Ref.[24] mainly comes from this choice of $\chi$ and accordingly a choice of an observer.
In \cite{Guedens:2011dy}, the authors chose $\chi$ so that it satisfies an approximate Killing equation. 
This may be one possibility of a thermodynamic interpretation of the Noether charge identity, but the four velocity of 
 the observer with the approximate Killing vector is not always 
parallel to the null generator of the null hypersurface, and the relation between the local Rindler horizon and the choice of an observer for whom the thermodynamic interpretation is applied is not very clear.
On the contrary, we chose the observer so that its four velocity becomes parallel to the generator of the null hypersurface. Hence the observer ($\chi$) is always directly related with the null hypersurface.
But, in compensation for that, we need to consider the effect of local change in the scaling of an observer's time.
Accordingly, we need to change the local scale of temperature. 
It is the reason why we needed to rescale the ``entropy" density as in (\ref{LchiNcharge}) or (\ref{entropychange}). 

We have seen that a thermodynamic formulation of the Einstein equation is observer dependent.
Because of the equivalence principle, such a property might  be inevitable 
if there exists a thermodynamical interpretation of space-time.
This kind of idea was emphasized in \cite{Padmanabhan:2009jb},
based on the fact that the concept of the horizon entropy or temperature 
in space-time cannot exist in the flat Minkowski space-time unless a Rindler observer is concerned.
The same will be applied to other thermodynamic quantities like the entropy production  $\delta N$. 
Furthermore, as mentioned in \cite{Hayward:2000fi}, 
the Noether charge relation (\ref{pre_Clausius}) can be interpreted as the law of an energy conservation
where the gravitational energy depends on the choice of observers. 
Hence it is not unlikely that the thermodynamic interpretation of space-time in (\ref{pre_Clausius}) 
is different depending on a choice of observers.
We hope to come back to this issue in future.

Finally we comment on yet another derivation of the Einstein equation from a different type of the Clausius relation,
which is reminiscent of the heat transfer equation on membrane of a black hole \cite{Parikh:1997ma}.
In the Einstein theory of gravity, the entropy is assumed to be proportional 
to the area of $(n-2)$-dimensional section ${\cal P}$ of a null hypersurface ${\cal H}$; $S \equiv  (\sqrt{\gamma}/4) d^{n-2}y $.
We take the parameter $\tau$ so that the tangent null vector on the hypersurface ${\cal H}$,
$k ^\mu=(\partial/\partial \tau)^\mu$, satisfies $(k \cdot \nabla) k ^\mu =\kappa k ^\mu$.
Then, by using the Raychaudhuri equation, it is easy to show the relation
\begin{align}
\frac{\kappa}{2 \pi} \left( \frac{d S}{d \tau} - \frac{1}{\kappa} \frac{d^2 S}{d \tau^2} \right) = \frac{1}{8\pi} \left( -\frac{n-3}{n-2} \theta^2 + \sigma^2 +R_{\mu\nu} k^\mu k^\nu \right) \sqrt{\gamma} d^{n-2}y. \label{raych}
\end{align}
The RHS of (\ref{raych}) is interpreted as an entropy production and a heat into the horizon (under the Einstein equation);
\begin{align}
\frac{\kappa}{2 \pi} \left( \frac{d S}{d \tau} - \frac{1}{\kappa} \frac{d^2 S}{d \tau^2} \right)=\delta N + \delta Q \ . \label{energydiss}
\end{align}
The second term in the LHS may be considered as a relativistic correction to the naive diffusion equation.
If we require the ``Clausius relation" (\ref{energydiss}) to hold,
we can generate the Einstein equation. 
This relation again explains why the original derivation of the Einstein equation of state needed an opposite sign of $\delta Q$.
The second term in the LHS of (\ref{energydiss}) is written as $d^{2}S/d\tau^{2}= (\dot{\theta}+\theta^2)S$, and the $\dot{\theta}$ term gives $\delta Q$ with a positive sign.
On the contrary, if we did not include the second term and simply expanded $dS/d\tau$ in $\tau$ as $dS/d\tau =\theta S =(\theta S)|_{\tau =0} + \tau (\dot{\theta}S+\theta^2 S)|_{\tau =0}+{\cal O}(\tau^2) $,
we would get an opposite sign of $\dot{\theta}$ and the sign of $\delta Q$ must be reversed in order to generate the Einstein equation.
This corresponds to the original derivation of the Einstein equation.
Since the LHS of (\ref{energydiss}) is nothing but $T \delta S$ in (\ref{entropychange}), 
we may shed a light on the  observer dependence  of  the entropy change in  (\ref{entropychange})
as the relativistic correction to the diffusion equation.

\section*{Acknowledgment}

\hspace{16pt}The research by S.I. is supported in part by Grant-in-Aid for Scientific Research (19540316) from MEXT, Japan.
The work of S.O. is partially supported by Grant-in-Aid for JSPS Fellows.
We are also supported in part by ``the Center for the Promotion of Integrated Sciences (CPIS)" of Sokendai.


\end{document}